\documentclass[aps,prl,twocolumn,superscriptaddress]{revtex4}

\usepackage[dvipsnames]{xcolor}

\usepackage[factor=1100]{microtype}     

\usepackage[font=sf, objectset=centering]{floatrow}
\usepackage{graphicx}

\usepackage{mathtools}                  
\usepackage{amssymb}                    
\usepackage{amsmath}                    
	
\usepackage{nicefrac}                   
\usepackage[
	mode=match,
	parse-numbers=false,
	parse-units=false,
]{siunitx}        
\usepackage{braket}                     
\usepackage{diffcoeff}                  
\usepackage{pdfpages}

\begin{document}

\title{Shocking a Shock Wave for Nonlinear Summation of GPa Pressures}

\author{Jet Lem}
\affiliation{Department of Chemistry, Massachusetts Institute of Technology, Cambridge, MA 02139, USA}
\affiliation{Institute for Soldier Nanotechnologies, Massachusetts Institute of Technology, Cambridge, MA 02139, USA}
\author{Yun Kai}
\affiliation{Department of Chemistry, Massachusetts Institute of Technology, Cambridge, MA 02139, USA}
\author{Maxime Vassaux}
\affiliation{Institut de Physique de Rennes, UMR CNRS 6251, Université de Rennes, 35042 Rennes, France}
\author{Steven E. Kooi}
\affiliation{Institute for Soldier Nanotechnologies, Massachusetts Institute of Technology, Cambridge, MA 02139, USA}
\author{Keith A. Nelson}
\email[Corresponding author,
\,\,]{kanelson@mit.edu}
\affiliation{Department of Chemistry, Massachusetts Institute of Technology, Cambridge, MA 02139, USA}
\affiliation{Institute for Soldier Nanotechnologies, Massachusetts Institute of Technology, Cambridge, MA 02139, USA}
\author{Thomas Pezeril}
\email[Corresponding author,
\,\,]{thomas.pezeril@cnrs.fr}
\affiliation{Department of Chemistry, Massachusetts Institute of Technology, Cambridge, MA 02139, USA}
\affiliation{Institut de Physique de Rennes, UMR CNRS 6251, Université de Rennes, 35042 Rennes, France}

\date{\today}

\begin{abstract}

Exploring shock-shock interactions has been limited by experimental constraints, particularly in laser-induced shock experiments due to specialized equipment requirements. Herein, we introduce a tabletop approach to systematically investigate the excitation and superposition of dual laser-induced shock waves in water. Utilizing two laser pulses, spatio-temporally separated and focused into a confined water layer, we identify the optimal superposition leading to the highest combined shock pressure. Our results demonstrate that combining two shock waves each of $\sim$0.6~GPa pressure yields an overall shock pressure of $\sim$3~GPa. Our findings, suggesting an inherent nonlinear summation from the laser excitation process itself and highlights a new pathway for energy-efficient laser shock wave excitation. 

\end{abstract}

\maketitle

Fundamental understanding of shock waves is of great importance in a range of studies, including planetary impacts, inertial confinement fusion, shock-induced fracture and spallation, primary traumatic brain injury, and shock-induced chemistry \cite{kraus2010shock,craxton2015direct,cottet1989spallation,veysset2019glass,arun2011studies,bradley1961shock}. As such, the experimental investigation of shock-shock interactions is of great interest. The use of multiple shock waves to push materials to exotics states of matter has been explored in the context of inertial confinement fusion as well as in “ring-up” shock experiments using conventional impact shock techniques \cite{craxton2015direct,brown2012microscopic,boehly2011multiple,kai2018formation}. However, the systematic study of the nonlinear superposition of shock waves remains largely unrealized.

Conventional shock wave experiments involve the acceleration of an impactor that, when incident on a sample surface, launches a shock wave through its depth; as in gas/powder gun experiments, micro-flyer plates, and split-Hopkinson bar type experiments \cite{brown2012simplified,field2004review}. Owing to the geometry of impact testing, such investigations are  limited to the study of individual planar shock waves. An alternative route towards shock wave experimentation, commonplace in the study of inertial confinement fusion and spallation, is direct laser-induced shock wave excitation \cite{craxton2015direct}. In these experiments, high-energy laser pulses are focused onto sacrificial ablator layers, pressurizing the contained materials. 

If, instead of an ablative layer, the sample itself is made to be optically absorptive, shock waves can be excited directly in the material of interest \cite{pezeril2011direct}. Absorption of the pump energy leads to ionization, plasma formation and expansion, launching a shock wave travelling in the sample. Such homogeneous direct laser-induced shock waves have been experimentally realized by several groups \cite{pezeril2011direct,gutierrez2021transient,radhakrishnan2022femtosecond,kai2017novel}. This experimental geometry benefits from the ability to spatially resolve the shock front travelling laterally in the sample plane, allowing imaging of the entire shock trajectory. It is well positioned to allow for shock wave excitation configurations previously unallowable by conventional impact and direct-drive experiments. One such geometry is a multiple-pulse scheme, for the study of shock wave interactions. 

Recent experiments by the Quinto-Su group were conducted involving transient time-delayed overlap of laser-induced shocks to investigate shock-shock interactions. They reported the observation of a nonlinear interaction caused by the collision of counter propagating direct laser-induced shock waves in water \cite{gutierrez2021transient}. Similar work by Radhakrishnan et al. investigated the densification of fused silica caused by the overlap of laser-induced shock waves \cite{radhakrishnan2022femtosecond}. The above two examples both involved the interference of counter propagating shock waves. Herein, we present experimental results regarding the overlap of co-propagating shock waves launched by direct laser excitation. We demonstrate that the superposition of shock waves allows for the nonlinear enhancement of achievable shock wave pressures.


\begin{figure*}
\includegraphics[width =\textwidth]{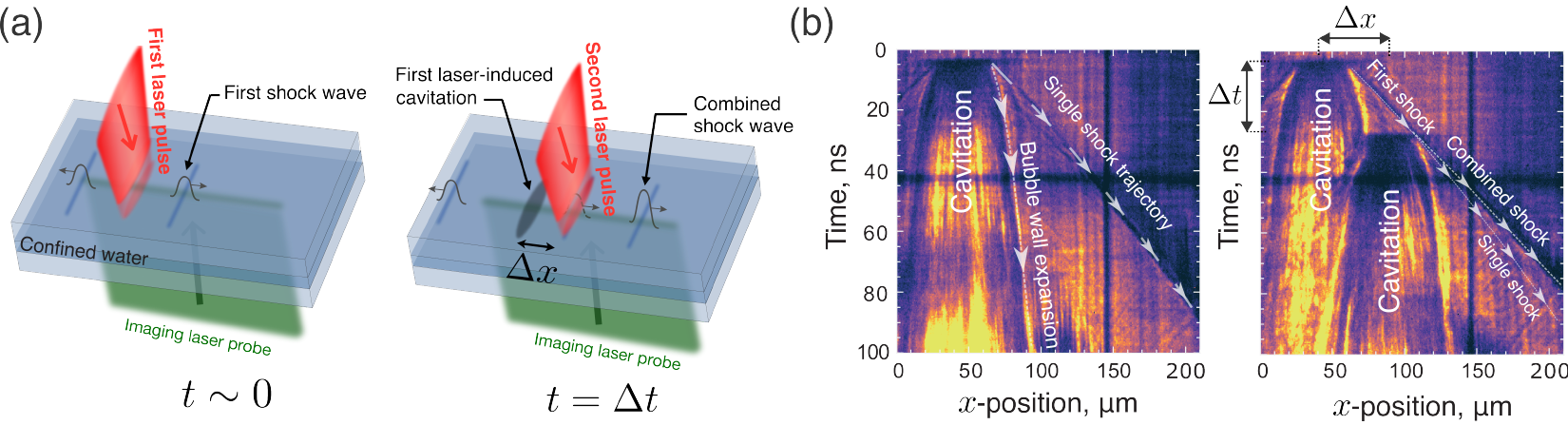}
\caption{\label{fig1} Dual laser-induced shock wave scheme (a) A first laser excitation pulse excites, in a confined water layer, two counter-propagating planar shock waves moving away from the excitation region. After time delay $\Delta t$, a second laser excitation pulse excites, at a laterally shifted position $\Delta x$, a second set of  shock waves. The superposition of the two shock waves in the co-propagating direction forms a combined shock wave. (b) Representative snapshot images recorded from the streak camera with a single laser pulse and with dual spatio-temporally spread laser pulses, each of these pulses with 1~mJ energy. The streak images, recorded in a single shot, are used to extract the full trajectory of the shock waves along the $x$-coordinate.}
\end{figure*}

Our laser-induced shock wave experiment has been described in depth previously \cite{pezeril2011direct,Veysset_2015,Veysset_2016,Veysset_2017,Veysset_2018,veysset2019glass}. In brief, a high energy laser pulse, delivered by the uncompressed output of a Ti:Sapphire amplifier (Coherent Legend Elite 800~nm -  150~ps), is focused as a line into the material of interest. Samples consisted of 25~\textmu m layers of water, doped with 5~wt\% carbon nanoparticles, sandwiched between two glass slides, each 100~\textmu m thick. The water layer thickness was set by aluminum spacers. Absorption of laser energy by the carbon nanoparticles causes vaporization of the water through flash heating, leading to a laser-induced cavitation bubble. Rapid expansion of the bubble walls leads to the generation of counter-propagating planar shock waves, travelling in the plane of the sample along the $x$-direction, see Fig~\ref{fig1}(a). A streak camera (Hamamatsu C4334) was used to acquire a 100~ns history of the shock event, in a single-shot, along the $x$-dimension perpendicular to the laser line of excitation, such that divergence of the planar shocks is negligible. Streak images were illuminated with a 150~ns, 532~nm imaging probe pulse (Coherent Evolution), that is spread along the entrance slit of the streak camera. A portion of the illumination pulse was diverted to a CCD camera for sample positioning.

For the shock superposition experiments, the laser-excitation pulse was split into two pulses, with a 25~ns delay between each pulse. An inter-pulse time $\Delta t$ of 25~ns was chosen to allow for sufficient resolution to image both excited shock waves, while avoiding dissipation effects on the first shock wave. The first laser-excitation pulse generate a shock wave as described above. The second laser-excitation pulse excites a shock wave which, at an optimum spatial position, overlaps and merges with the first co-propagating shock wave. Example of streak images are shown in Fig~\ref{fig1}(b). As seen in these images, the laser-excitation of the confined water layer leads to vaporization, cavitation, and bubble expansion, giving rise to shock excitation. The subsequent shock propagation appears on the images as a straight line, see for instance the single shock trajectory on the left side of Fig~\ref{fig1}(b). The slope of the shock trajectory at each $x$-time- coordinate is the instantaneous shock speed $U_s$. The shock wave speed is related to the shock wave pressure $P$, through Eq.~(\ref{eq:1}), where acoustic wave speed $c_0$= 1.45 km/s, density $\rho_0$= 1.000 g/cm$^3$ for water, and the denominator is an empirical coefficient for water \cite{gojani2016shock}. In case of the combined shock in Fig~\ref{fig1}(b), there is an observable change in slope of the combined shock, when compared with the single shock case, indicating an increase in the shock wave pressure in the case of the combined shock.
\begin{equation}
P = \rho_0 \, U_s \frac{(Us - c_0)}{1.78} \,\, \text{[GPa]}    \label{eq:1}
\end{equation}


\begin{figure}[!htb]
\includegraphics[width =\textwidth]{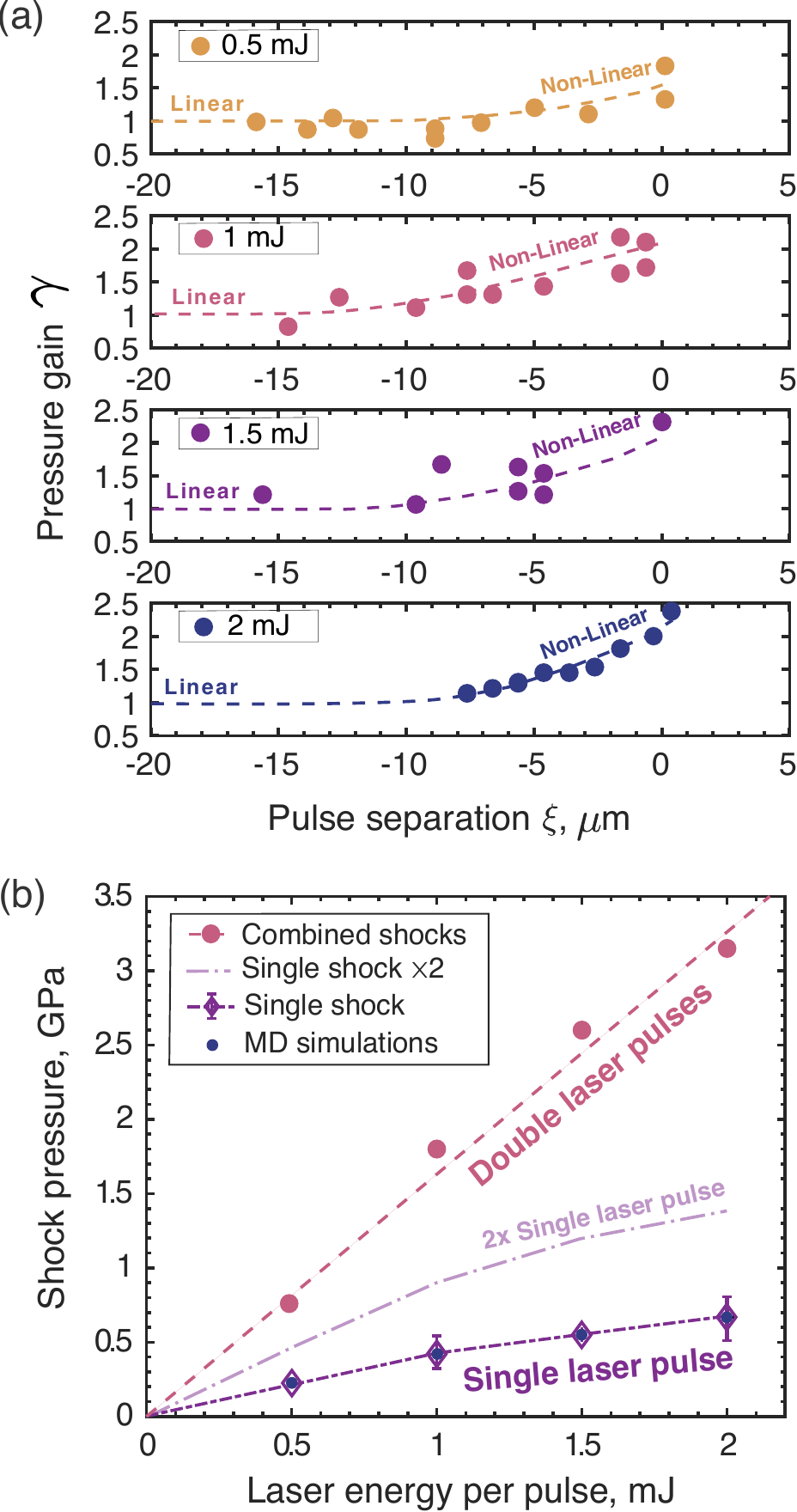}
\caption{\label{Fig2} (a) Calculated pressure gain $\gamma$ for the superposition of two laser generated shock waves as a function of the excitation pulse spacing $\xi$. A pressure gain of 1 corresponds to linear superposition of the pressure of each individual shock. At any given energy tested, the maximum observed pressure gain corresponds to perfect spatio-temporal overlap $\xi$~=~0. (b) Extracted experimental shock pressures of a single shock and combined shocks, plotted against input excitation energy. Note that the laser energies refers to the energy per pulse. For example, the combined shock point at 2~mJ was excited with a total input pulse energy of 4~mJ.}
\end{figure}

To quantify the effect of the shock wave superposition, we first determine the pressure excited by a single laser pulse at four laser pulse energies (0.5, 1.0, 1.5, and 2.0 mJ). We will refer to these values, plotted in Fig~\ref{Fig2}(b), as $P_{single}$. We then conducted measurements with two excitation pulses, tuning the spatial separation between the two excitations $\Delta x$ to optimize the shock superposition. The shock wave pressure resulting from the combination of two waves will be referred to as $P_{double}$. To quantify the pressure enhancement allowed by shock wave superposition, we introduce two parameters, the pressure gain parameter $\gamma  = P_{double}/(2\times P_{single})$ and a normalized spatial coordinate $\xi$. A pressure gain of $\gamma = 1$ indicates the pressure of the combined shock is equal to $P_{double} = 2 \times P_{single}$. We will refer to $\gamma = 1$ values as the linear superposition regime, labeled in Fig~\ref{Fig2}(a). The normalized spatial coordinate $\xi$ is defined below. $$\xi =  \Delta x - U_s \times \Delta t \,\, $$ A spatial coordinate value of 0 indicates that the second laser pulse is incident on the sample as the first shock is propagating through that region. In other words, the second laser pulse is focused onto the first propagating shock wave. A negative $\xi$ value indicates that the second laser pulse is focused behind the first shock wave. A positive $\xi$ value, corresponds to the case for which the first shock wave is behind the second shock wave. In the case of positive $\xi$, the first shock travels through the cavitation zone of the second excitation and will be ignored in this investigation. 

Fig~\ref{Fig2}(a) plots the pressure gain $\gamma$ for 0.5~mJ, 1~mJ, 1.5~mJ, and 2~mJ input laser energies, per excitation pulse, at different spatial coordinates $\xi$. At each energy tested, $\gamma$ reaches a maximum value of nearly 2 as $\xi$ approaches 0. This indicates a nonlinear increase in pressure caused by the superposition of the shock waves. We observe that the region along the $\xi$-axis for which nonlinear superposition occurs narrows with increasing laser pulse energy. This is a result from an increase in the shock speed with laser pulse energy, altering the time that the first shock wave remains in the second laser-excitation region.

The maximum pressures achieved through shock wave superposition, for each laser pulse energy, are presented in Fig~\ref{Fig2}(b). Across all tested energies, the combined shock waves consistently reached higher pressures than could be reached using single laser pulses with equivalent total input laser fluence. Comparing the linear superposition line, representing values equal to $2\times P_{single}$, with the peak pressures achieved from the superposition of two shock waves, there is a clear disparity. In terms of shock excitation efficiency, the data presented in Fig~\ref{Fig2}(b) reveal a remarkable advantage in the superposition of two shock waves compared to a single shock wave. For instance, at a laser energy of 2~mJ for a single pulse, the pressure achieved is approximately 0.7~GPa. However, when the energy is split between two excitation pulses of 1~mJ each, the combined shock generates a pressure of around 1.8~GPa, more than double the pressure produced by a single pulse. This efficiency gain is consistent across various laser energies, with the benefit of splitting the laser beam into two spatio-temporally overlapping pulses becoming more pronounced at higher energies. At lower energies, for example, a 1~mJ pulse produces 0.5~GPa, while two 0.5~mJ pulses yield 0.8~GPa.

Additionally, the combined shock waves appear to follow a linear trend with an excitation efficiency of 0.8~GPa/mJ. This is opposed to the single shock excitation efficiency, which appears to plateau between 1-2~mJ. This plateau may be attributed to plasma formation, saturation of carbon nano-particle absorption, bubble expansion, or other such effects that would mitigate laser energy conversion at higher input energies. This is a well-documented drawback of laser-excitation of shock waves \cite{Berthe, pezeril2011direct, Metalens}. The results presented in Fig.~\ref{Fig2}(b) demonstrate that by separating the excitation into lower energy pulses, these drawback may be avoided.

\begin{figure}[!htb]
\includegraphics[width =\textwidth]{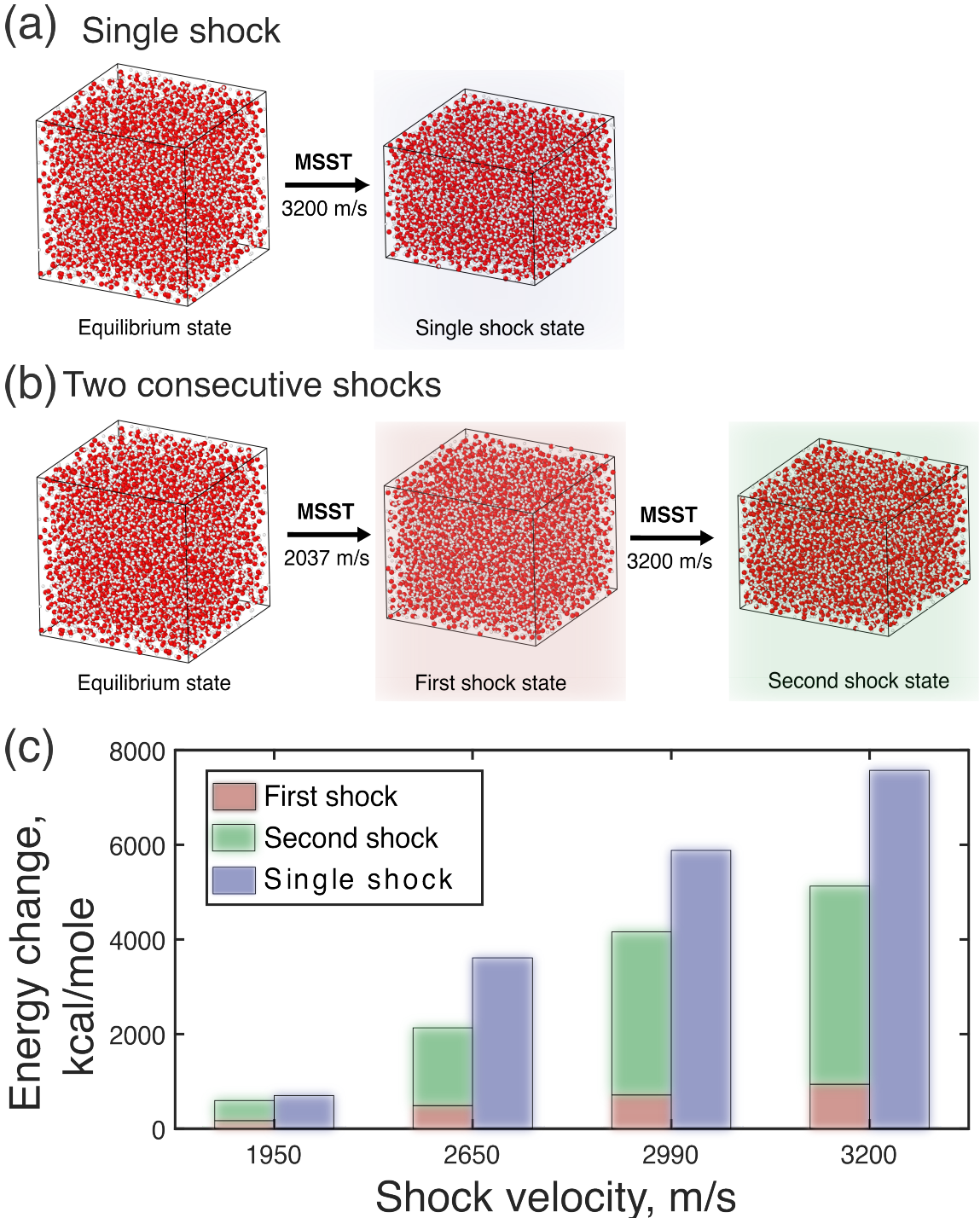}
\caption{\label{fig_simul} Shock simulation of the molecular water model at equilibrium and in shocked states after (a) a single-step shock excitation or (b) two-step consecutive excitation. The water molecules are all shown in this example for the simulation of a final shock wave velocity of $3200$ m/s. (c) Internal energy increase required to induce shock waves at $1950$, $2650$, $2990$ and $3200$ m/s with a single (right bar) or two consecutive (left bar) shocks in the molecular model.}
\end{figure}

To better understand the underlying reasons for the efficiency gain, we conducted molecular dynamics simulations using the Multi-Scale Shock Technique (MSST) to model the behavior of a molecular system subjected to either a single or two superimposed shock waves \cite{reed2003method}, as illustrated in Fig.~\ref{fig_simul}(a). The MSST allows for precise control over the dimensions of a molecular model over time to induce a shock wave at a specified shock velocity. In our simulations, we modeled the dynamics of a periodic system containing $4,000$ water molecules (bulk water) under shock conditions for 0.5~ns, utilizing the TIP4P interatomic potential. This approach enables us to calculate the internal energy increase in the molecular system in response to the induced shock state, see Fig.~\ref{fig_simul}(b). Further details on the implementation of the molecular model, including validation against equilibrium data (structure, density) and single-shock experimental results (equation of state), are provided in the supplementary material.

We focus on the thermodynamic response of water molecules subjected to shock waves at target velocities of $1950$, $2650$, $2990$, and $3200$~m/s, which correspond to the maximum measured shock velocities for the shock wave superposition experiments conducted at 0.5, 1.0, 1.5, and 2.0~mJ laser pulse energies, respectively (see combined shock data in Fig~\ref{Fig2}(b)). To evaluate the enhancement afforded by separating the shock wave excitation into two pulses, we simulate and compare two scenarios: a single-step shock excitation and a two-step shock excitation. In the single-step scenario, a shock wave is launched at the final target velocity and the corresponding internal energy change is calculated. 

In the two-step shock scenario, a weaker shock wave, below the final target velocity, is first initiated at velocities of $1664$, $1856$, $1953$, and $2037$~m/s, corresponding to the velocities experimentally measured for single laser pulse excitations in water at 0.5, 1.0, 1.5, and 2.0~mJ laser pulse energies, respectively (single shock data in Fig~\ref{Fig2}(b)). A second shock wave is then introduced into the system, with the energy selected to ensure the final target velocity is reached. We calculate the total internal energy for all water molecules at each stage of the simulation. 

Figure \ref{fig_simul}(b) compares the calculated internal energy changes for two scenarios outlined above. The data reveals that to reach the same final thermodynamic state, the two-step excitation requires less internal energy change than a single-step shock excitation (Fig.~\ref{fig_simul}(b). Additionally, this disparity becomes more pronounced as the target shock wave velocity increases. This indicates that, in the consecutive shock scenario, the initial thermodynamic state change induced by the first shock lowers the energy barrier necessary to attain the final shocked state. 

The numerical modeling aligns with the experimental results, demonstrating that, in terms of energy efficiency, it is beneficial to split the input laser pulse energy into two pulses for more efficient shock excitation. However, the simulations indicate at best a 20-30\% reduction in energy input for the dual-shock scenario compared to the single-shock case. Experimentally, however, the improvement in shock efficiency with combined shocks is far more pronounced, as evidenced from the comparison between the combined shocks and the linear superposition in Fig~\ref{Fig2}(b). This discrepancy suggests that rather than the thermodynamics of the system, the significant nonlinear efficiency gain originates primarily from the laser-excitation process itself, particularly the highly nonlinear laser-induced cavitation process \cite{Veysset_2018} that is not taken into account in the MSST model.

\begin{figure}[!htb]
\includegraphics[width =\textwidth]{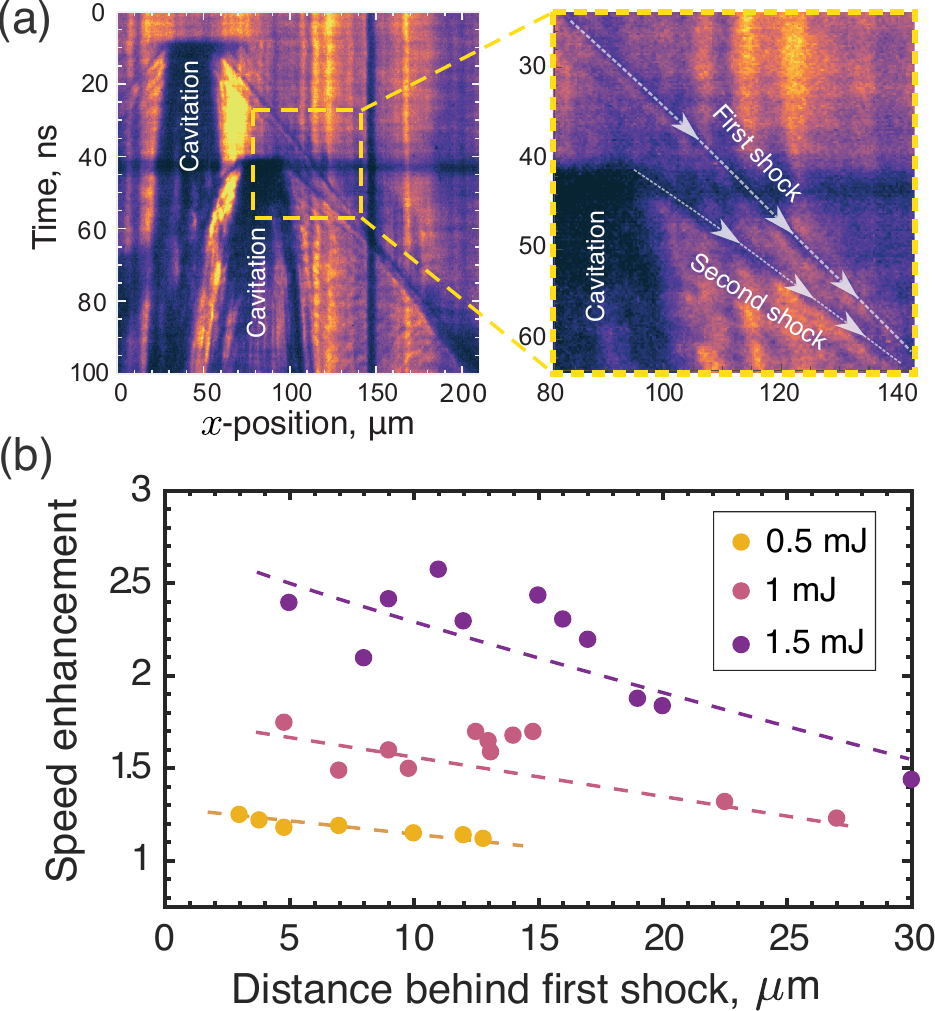}
\caption{\label{Fig4}(a) Representative image recorded from a second shock wave excited well-behind the first shock wave. The speed enhancement of the second shock is evident in the inset, and represented by arrows outlining the trajectory. (b) Speed enhancement of a trailing shock wave as a function of excitation position behind the first shock. Trailing shock waves demonstrate speed enhancement up to tens of microns behind the initial shock, indicative of lasting material densification after shock excitation.}
\end{figure}

Additional experiments were conducted where the spatial separation, $\xi$, between the two laser pulses was deliberately detuned, such as to observe the behavior of the second shock travelling in the wake of the first shock. In these experiments, the first shock wave was excited at 0.5~mJ, 1.0~mJ, and 1.5~mJ, laser pulse energies, followed by a second laser excitation at a fixed energy of 0.5~mJ per pulse, arriving at the sample 25~ns after the first excitation. The results show that the second shock exhibits significantly enhanced speeds for trailing distances of up to tens of microns behind the initial shock. This speed increase suggests that the water layer remains in a post-shock, densified state up to tens of microns behind the propagating shock front. The effect is summarized in Fig.~\ref{Fig4}(b), which shows the shock speed enhancement—calculated as the ratio of the trailing shock speed to the single shock speed at 0.5 mJ laser pulse energy. This experimental technique may offer valuable insights into material densification induced by shock waves. Specifically, if instead of a second shock wave, which perturbs the material properties, an acoustic wave with negligible pressure loading were excited, it could allow for direct measurement of changes in the longitudinal acoustic wave speed as a function of distance behind the shock. Unfortunately, this investigation was not feasible with the current setup due to the lack of sensitivity in detecting small-amplitude acoustic waves. To achieve this, a more sensitive probing method, such as optical interferometry, would need to be implemented. This "catch-up" behavior of the second shock is consistent with previous findings \cite{kai2018formation}, where it was observed that trailing shocks will eventually catch the front-running shock.


In this study, we experimentally demonstrate that shock wave superposition is an effective method for significantly amplifying experimentally achievable shock wave pressures. The most substantial pressure increase occurred when the spatial separation $\xi$ was zero, indicating perfect overlap of the two shock waves, with the second laser pulse directly exciting the first. Under these conditions, the superposition of two 0.6~GPa shock waves produced pressures as high as 3~GPa. This nonlinear summation emphasizes the increased efficiency of shock wave generation via multiple laser pulse irradiations. Though molecular dynamics simulations revealed an intrinsic thermodynamic efficiency gain allowed by a two-step excitation, the magnitude of this gain was small compared to that observed experimentally. Since the molecular dynamics simulation does not fully account for the intricacies of the laser excitation, particularly the highly nonlinear, laser-induced cavitation effect, our results suggest that this nonlinear efficiency gain primarily stems from the complex dynamics inherent in the laser-excitation process.

The combined shock waves exhibited an excitation efficiency of 0.8~GPa/mJ, following a linear trend along the energies tested, unlike the single-shock excitation, which plateaus significantly. We suspect this plateau is due to factors such as plasma formation, saturation of carbon nanoparticle absorption, or other effects that impede efficient laser energy conversion. By contrast, shock wave superposition presents a potential pathway to achieve higher laser-induced pressures while mitigating these parasitic effects. Further enhancement in excitation efficiency could be achieved by extending the technique beyond two excitations, potentially involving multiple excitations, drawing inspiration from results obtained in the non-destructive regime using an optical scheme for spatio-temporal superposition of more than 20 laser beams \cite{Zebra}. Additional improvements could be realized by employing multiple cylindrically converging shock waves \cite{Metalens,patent,gutierrez2022bullseye}. This approach, particularly when applied with multiple waves rather than just a single one, offers a promising avenue to significantly boost the laser excitation efficiency of shock waves.

\section{Acknowledgments}
 We gratefully acknowledge Fabio De Colle and Pedro Quinto-Su from Instituto de Ciencias Nucleares, Universidad Nacional Autónoma de México for extremely valuable scientific discussions. This research was supported by the U. S. Army Research Office under Cooperative Agreements Number W911NF-22-2-0170 and W911NF-18-2-0048, the Basic Research Office, Office of Under Secretary of Defense for Research and Engineering (OUSD R\&E) under the Laboratory University Collaboration Initiative (LUCI) program, as well as from R\'egion Bretagne under SAD grant CHOCONDE and from Rennes m\'etropole.

\bibliography{Main}

\clearpage


\onecolumngrid  
\thispagestyle{empty}  
\noindent
\centering


\section*{Supplementary material}

# Shocking a Shock Wave for Nonlinear Summation of GPa Pressures

Jet Lem,[1,2] Yun Kai,[1] Maxime Vassaux,[3] Steven E. Kooi,[2] Keith A. Nelson,[1,2] and Thomas Pezeril[1,3,*]

[1]*Department of Chemistry, Massachusetts Institute of Technology, Cambridge, MA 02139, USA*

[2]*Institute for Soldier Nanotechnologies, Massachusetts Institute of Technology, Cambridge, MA 02139, USA*

[3]*Institut de Physique de Rennes, UMR CNRS 6251, Université de Rennes, 35042 Rennes, France*

## MOLECULAR DYNAMICS SIMULATIONS

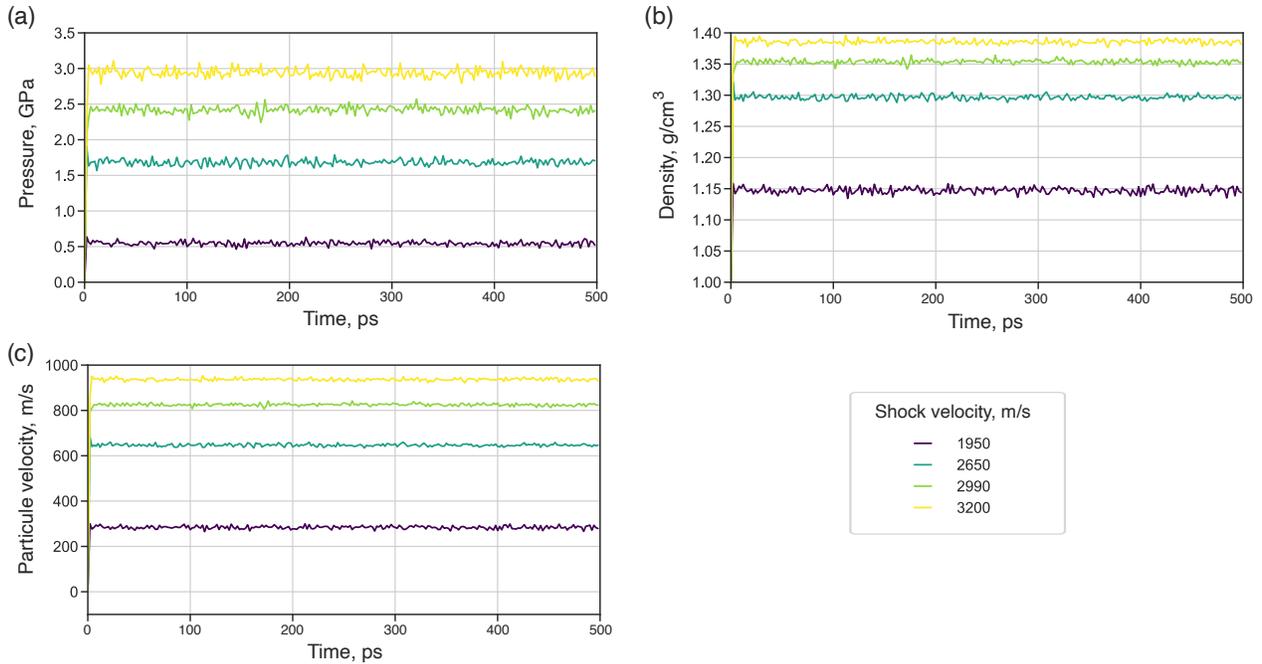

FIG. S1. Evolution of the pressure, the particle velocity and the density during 0.5 ns molecular dynamics simulations of single shocks at velocities of 1950, 2650, 2990 and 3200 m/s using the MSST.

We perform molecular dynamics simulations of a molecular model of water using LAMMPS [1]. Our molecular model consists in 4000 water molecules enclosed in a periodic box. We employ the TIP3P interatomic potential for water featuring rigid bond interactions [2]. We rely on the validated procedure established by Neogi et al. [3] which uses the Multi-Scale Shock Technique (MSST). The technique updates positions and velocities of the atoms in the system at each timestep to reproduce the compression induced in the system when trapped in the shock wave. The MSST input is the shock wave velocity to be simulated. The technique outputs associated particle velocity. In addition, we are the able to compute pressure, density and energy changes associated with the induced compression in the system. We set the parameters $q$ and $t_{scale}$ at 35 and 0.02 respectively, such as to observe steady compressive shock waves during the course of the simulations (see figure S1).

We first verify our model on the predicted density at equilibrium, we find a value of 0.98 g/cm³ which is consistent with simulations relying on the TIP3P interatomic potential. We then validate the approach for our molecular model of water by computing pressure versus particle velocity plots at the shock wave velocities observed experimentally. We perform such validation on single shock experiments in order to compare with the known and well-established equation of state of water [4]: $P = \rho_0(c_0 + S * u_p) * u_p$ with $P$ the pressure, $u_p$ the particle velocity, $\rho_0$ the density of water in ambient conditions and the parameters $c_0 = 1.647$ km/s and $S = 1.921$. The validation can be found in figure S2.

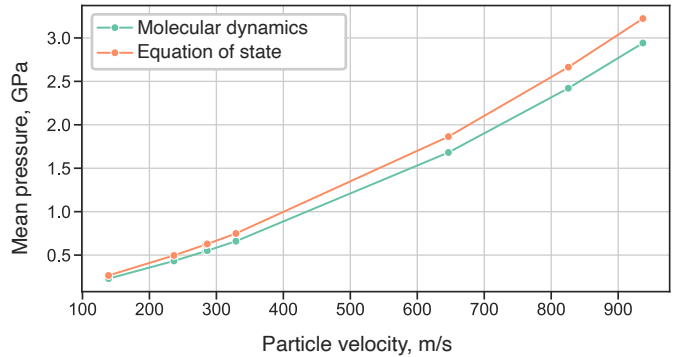

FIG. S2. Evolution of the shock pressure versus particle velocity for single shock scenarios: comparison between the equation of state of water [4] and the prediction from our molecular dynamics simulations.

## SET UP

**Pump path.** The pump laser used in the experiments is derived from the uncompressed output of a 1 kHz Ti:Sapphire regenerative amplifier (Coherent Legend) with a central wavelength of 800 nm, a pulse duration of 300 ps FWHM and a maximum energy per pulse of ¡4 mJ. The repetition rate of the laser is decreased down to 10 Hz, in order to ensure that a mechanical shutter placed at the exit of the amplifier can select a single laser pump pulse. This is a requirement for the single-shot experiments that we have conducted. A half-wave plate and polarizing beam splitter combination placed after the shutter acts as a variable attenuator to tune the overall pump energy injected into the pump path of the setup. Afterward, the beam is split into two separate beams using another half-wave plate and polarizing beam splitter combination. One beam is directed to the sample immediately, while the second beam is delayed by 25 ns by extending its beam path using multiple mirrors. Both beams are spatially recombined using a polarizing beam splitter before being focused onto the sample through a 50 cm cylindrical lens and a ×10 microscope objective (long working distance Mitutoyo), as illustrated in Fig.S3. This configuration creates a line-shaped focus of $5\mu\text{m}\times250\mu\text{m}$ FWHM dimensions on the sample surface.

**Imaging Probe.** The imaging probe beam is derived from the Nd:YAG pulsed beam used to pump the Ti:Sapphire crystal for amplification. The imaging probe beam operates at 100 Hz, at a central wavelength of 532 nm, with an average power at the sample lower than 10 mW. The imaging probe beam goes through a 30 cm lens, is reflected from a dichroic mirror to mix the imaging probe beam with the pump beams that all go through the same ×10 microscope objective. The imaging probe beam is shaped as a spot on the sample surface. The imaging probe beam is then imaged with the conjunction of a second identical ×10 microscope objective and a 75 cm spherical lens on the entrance slit of the streak camera (Hamamatsu C4334), see Fig.S3. Note that the timing of the imaging probe is controled electronically in order to set the single-shot pump-probe delay. The imaging probe beam provides almost continuous illumination for capturing the shock event. For alignment purposes, a portion of the imaging probe beam is directed as well onto a CCD camera (Hamamatsu Orca Flash), enabling fast and precise positioning of the sample. Since the sample is locally damaged after each laser shot, the user translates the sample to a new, pristine area prior to each subsequent shot. Accurate repositioning and focus are ensured using the CCD image feedback, allowing for efficient sample preparation between laser shots.

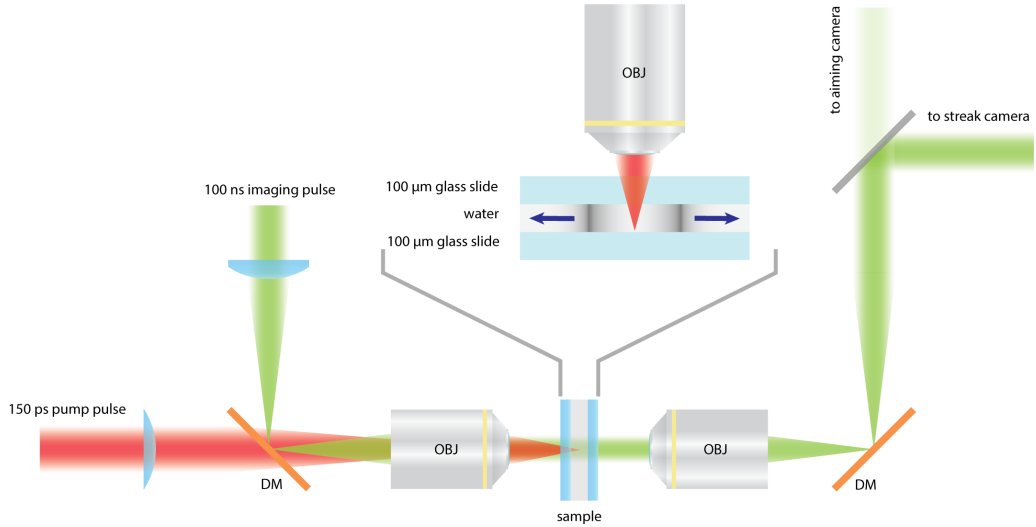

FIG. S3. Schematic illustration of the experimental setup. The pump pulses of 300 ps duration, 800 nm wavelength, are focused on the sample with both a cylindrical lens and a ×10 microscope objective (OBJ) to form a line-shaped beam at the sample. A 100 ns imaging pulse is directed to the same microscope objective after passing through a cylindrical lens with perpendicular orientation–as compared to the one on the pump bath. The imaging probe shaped as a thick line is imaged at the entrance of the slit of the streak camera. (DM) Dichroic mirrors used to reflect the 532 nm imaging probe and to transmit the 800 nm pump beams. .

## SET OF STREAK CAMERA DATA

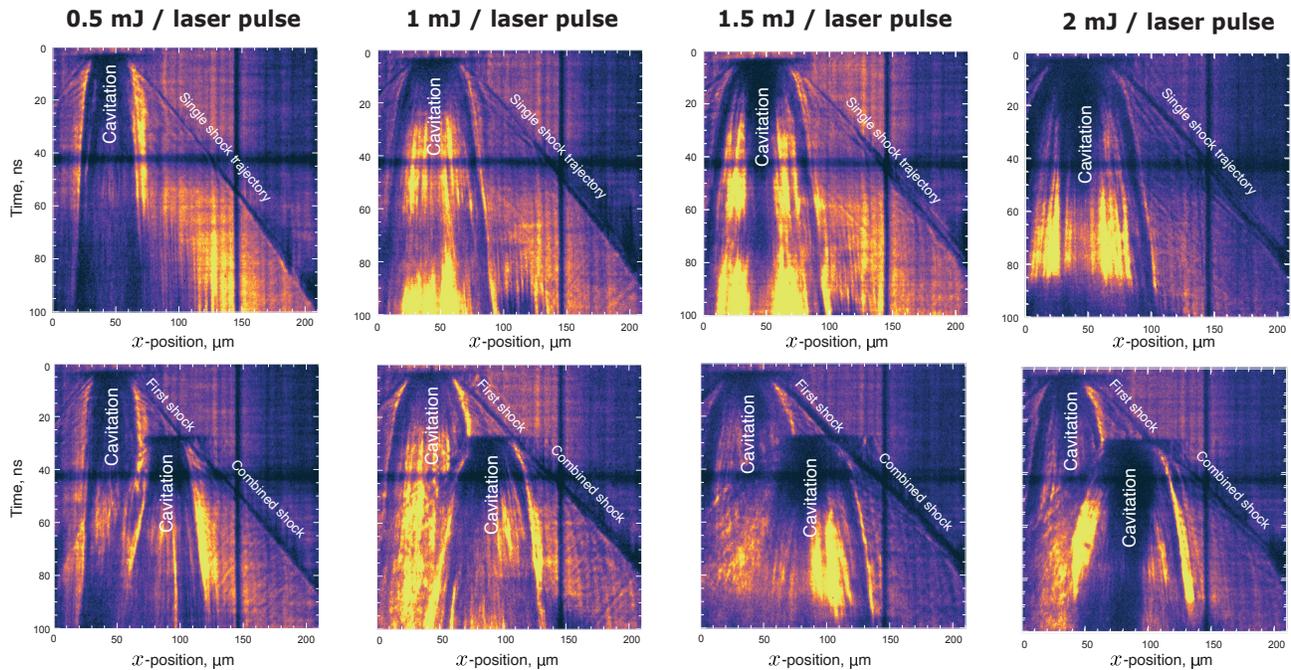

FIG. S4. Example streak camera data. Top row) Single line excitation of a shock wave in water with 5 wt% carbon nanoparticles. The laser pulse energies are labelled above the images. The stationary, gradually expanding dark region is the bubble that forms in the excitation region. The shock wave, labelled using a white arrow, is shown travelling away from the bubble. Bottom row) Example of double-shock excitation. The first shock wave travels for 25 ns before the second shock excitation happens. The combined shocks can be seen travelling away from the second shock excitation region.

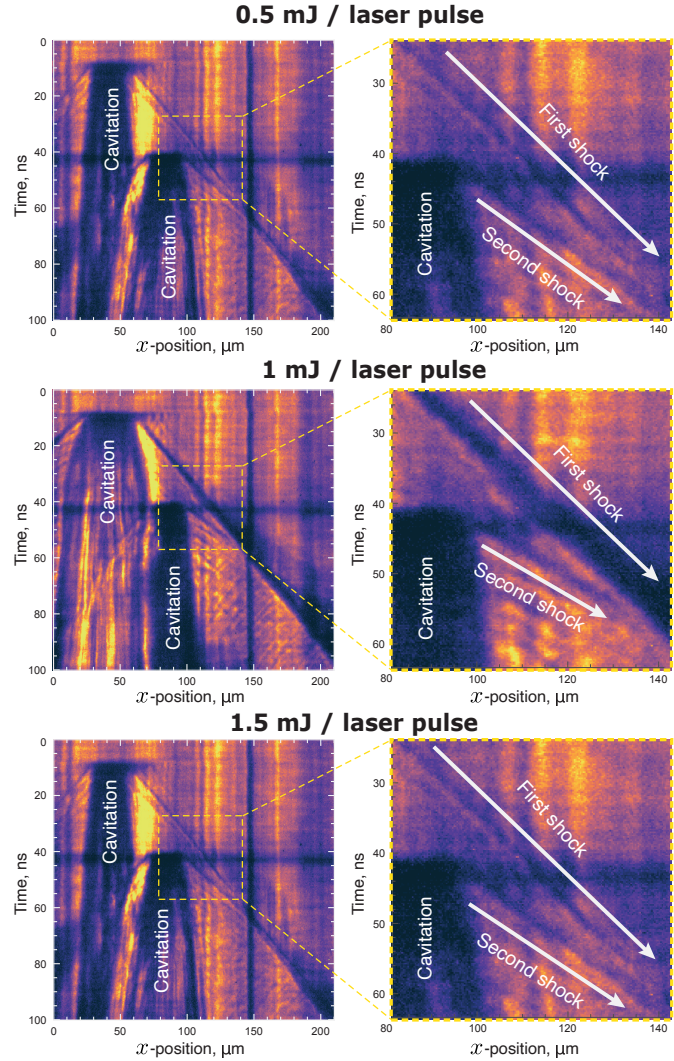

FIG. S5. Representative streak camera images for the trailing shock wave experiment. The sample is a confined water layer with 5 wt% carbon nanoparticles. The trailing shock is excited with a 0.5 mJ laser pulse energy for all tests. The laser pulse energies used to excite the first shock wave are labelled above the images. The right insets provide a zoomed view, highlighting the faster second (trailing) shock wave, which catches up with the first shock wave.

## FS-SNAPSHOT MEASUREMENTS

To test the universality of nonlinear superposition, we conducted single- and double-shock experiments using a different setup in a different laboratory, varying both laser and sample configurations. A Nd:YAG laser (Ekspla SL235, 1064 nm, 200 ps) was line-focused into a sample consisting of a 25 μm water layer doped with either 2.5 wt% or 5 wt% carbon nanoparticles (as specified in the corresponding figure caption). The water layer was sandwiched between two quartz slides (25.4 mm diameter, 200 μm thickness), with its thickness defined by aluminum spacers.

A CMOS camera (Hamamatsu Orca-Fusion) captured femtosecond snapshot images of the shock event, with exposure time controlled by a Ti:Sapphire amplifier (Coherent Libra, 800 nm, 200 fs) as the probe. An avalanche photodiode (Hamamatsu C5658), connected to an oscilloscope, recorded the arrival times of the pump and probe laser pulses on the water sample. From the fs-snapshot images presented in this supplementary section, we extracted the shock propagation distance as a function of time.

In summary, we confirm the observation of nonlinear shock enhancement across various configurations, including:

1. Water samples doped with varying ink concentrations. 2. Shocks generated using different laser parameters, such as energy, wavelength, pulse duration, and the delay between two pulses.

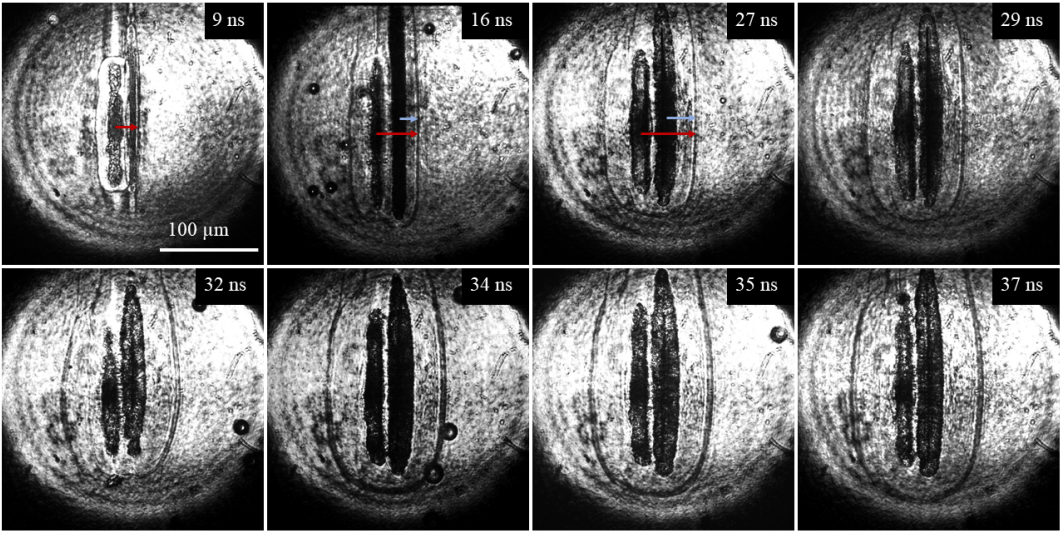

FIG. S6. Representative femtosecond snapshot images for double shock experiments. The sample was a 25 µm layer of water, doped with 2.5 wt% carbon nanoparticles. The time delay between the two laser lines was fixed at 8 ns. Each laser line was 2 mJ in energy. The phase-matching direction is from left to right in the images. The red and blue arrows indicate the propagation distance of the first and second shock, correspondingly.

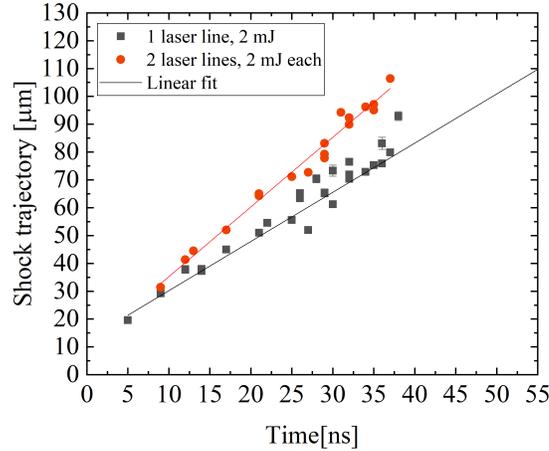

FIG. S7. Shock trajectories extracted from femtosecond snapshot images by measuring the propagation distance vs time stamp. The sample was a 25 µm layer of water, doped with 2.5 wt% carbon nanoparticles. The time delay between the two laser lines was fixed at 8 ns. The single line shock speed was 1850 ± 90 m/s and the pressure was 0.42 ± 0.11 GPa. The double line shock speed was 2503 ± 77 m/s and the was pressure 1.48 ± 0.15 GPa. Therefore, the pressure gain $\gamma$ was approximately $1.48/(2\times0.42) \simeq 1.8$.

| | 2 mJ experiments | 1 mJ experiments |
|---|---|---|
| Ink wt% in water sample | 2.5% | 5% |
| Shock, single laser line | 1850 m/s, 0.42 GPa | 1818 m/s, 0.38 GPa |
| Shock, two laser lines | 2503 m/s, 1.48 GPa | 2407 m/s, 1.29 GPa |
| Laser configuration | 8 ns delay between pulses | 14 ns delay between pulses |
| Pressure gain $\gamma$ | 1.8 | 1.7 |

TABLE S1. Key features of the fs-snapshot measurements.

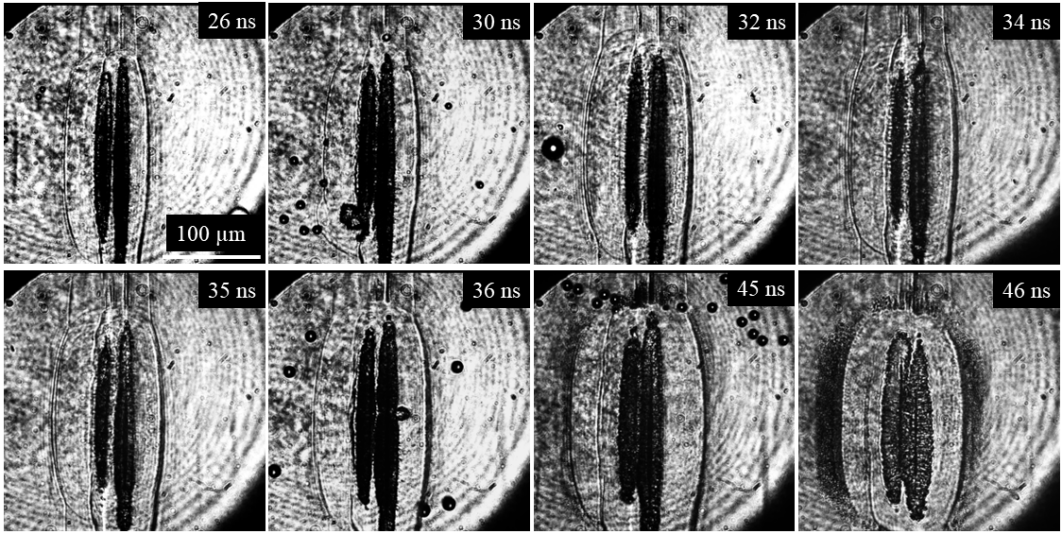

FIG. S8. Representative femtosecond snapshot images for double shock experiments. The sample was a 25 µm layer of water, doped with 5 wt% carbon nanoparticles. The time delay between the two laser lines was fixed at 14 ns. Each laser line was 1 mJ in energy.

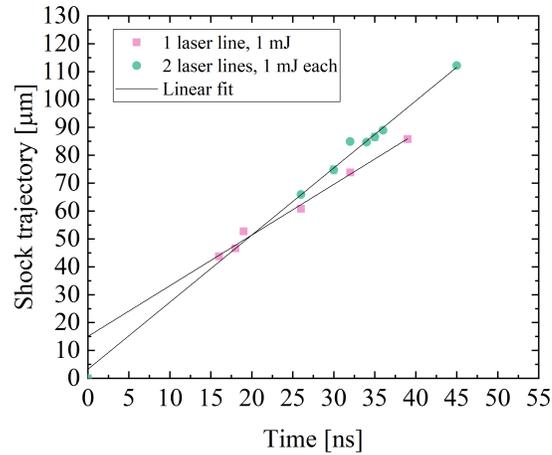

FIG. S9. Shock trajectories extracted from femtosecond snapshot images. The sample was a 25 µm layer of water, doped with 5 wt% carbon nanoparticles. The time delay between the two laser lines was fixed at 14 ns. The single line shock speed was 1818 ± 80 m/s and the pressure was 0.38 ± 0.10 GPa. Double line shock speed was 2407 ± 127 m/s and the pressure was 1.29 ± 0.24 GPa. Therefore, the pressure gain $\gamma$ was approximately 1.6.

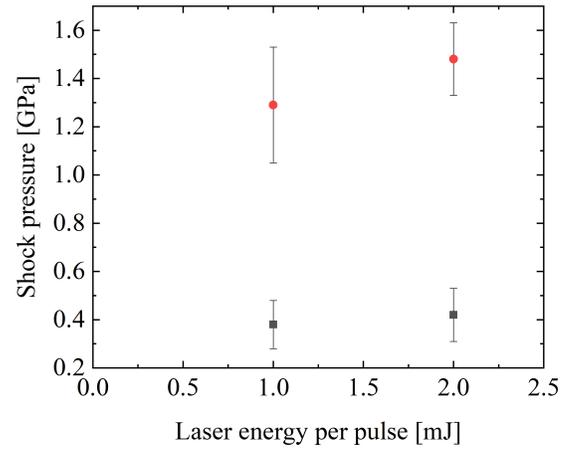

FIG. S10. Extracted experimental shock pressures of a single shock and combined shocks, plotted against input laser energy.

---



\end{document}